\documentstyle[11pt,aasms4]{article}

\def\mdotscale{\left({\dot m \over 0.1}\right)}
\def\Thetascale{\left({\Theta \over 0.1}\right)}
\def\xscale{\left({x \over 10}\right)}

\def\ana{A. \& A.}
\def\mnras{M.N.R.A.S.}
\def\apj{Ap.J.}
\def\apjl{Ap.J. Lett.}

\begin{document}
 
\title{A New Equilibrium for Accretion Disks Around Black Holes}
    
\author{J.H. Krolik\altaffilmark{1}}

\altaffiltext{1}{Department of Physics and Astronomy, Johns Hopkins
University, 3300 N. Charles St., Baltimore, MD 21218; jhk@pha.jhu.edu}
       
\begin{abstract}

      Accretion disks around black holes in which
the shear stress is proportional to the total pressure, the accretion rate
is more than a small fraction of Eddington, and the matter is distributed smoothly are both thermally and viscously unstable in their inner portions.
The nonlinear endstate of these
instabilities is uncertain.  Here a new {\it inhomogeneous}
equilibrium is proposed which is both thermally and viscously stable.
In this equilibrium the majority of the mass is
in dense clumps, while a minority reaches temperatures $\sim 10^9$~K.
The requirements of dynamical and thermal equilibrium completely determine
the parameters of this system, and these are found to be
in good agreement with the parameters derived from observations of accreting
black holes, both in active galactic nuclei and in stellar binary systems.

\end{abstract}


\section{Introduction}

    If the dissipation rate in an accretion disk is proportional
to the pressure (Shakura \& Sunyaev 1973), the disk is both viscously
(Lightman \& Eardley 1974) and thermally unstable (Shakura \& Sunyaev 1976)
wherever the radiation pressure exceeds the gas pressure.  Unfortunately,
this is exactly what happens in the inner parts of disks around black holes
whenever the accretion rate is greater than a small fraction of Eddington. 
Because the innermost radii are where most of the energy is released, these
instabilities may disrupt the most interesting portions of these disks.

     The ultimate result of these instabilities--that is, the actual
structure of disks occurring in Nature---remains unknown.
Shapiro, Lightman, \& Eardley (1976) suggested that the endpoint is
a two-temperature fluid in which the ions are much hotter than the electrons,
with the contrast maintained by the inefficiency of heat transfer by
Coulomb collisions.  However, this equilibrium is also thermally unstable
(Piran 1978).  Ichimaru (1977), Rees et al. (1982), and Narayan \& Yi (1995)
have argued that the two-temperature solution can be stabilized by
the inward advection of heat.  However, ions are hotter than electrons
only if most of the dissipated heat is given to the ions, a much-disputed point
(e.g. Bisnovatyi-Kogan and Lovelace 1997; Quataert 1997; Blackman 1997;
Gruzinov 1997).  In addition, there may be particle-wave interactions
which greatly accelerate ion-electron heat transfer (Begelman \& Chiueh 1987).
Another possible stable equilibrium is one in which the dissipation is
concentrated in a disk corona (Svensson \& Zdziarski 1994). 
However, there is no known mechanism that can take most of the accretion
energy liberated inside the body of the disk, and transport it with only
minimal losses to its surface.

     The thermal instability (whose growth rate is larger than that of the
viscous instability by $\sim (r/h)^2$, for disk thickness $h$ at radius
$r$) is driven by a feedback loop in which increased radiation pressure causes
increased dissipation, which then, because the surface density, and
hence the optical depth, are fixed on the thermal timescale, leads to
greater radiation pressure.  If increasing pressure caused the effective
optical depth to diminish, the instability could be quenched.  Clumping
of the disk matter would lead to just this result, for photons
find their way out through the most transparent channels.

     In this {\it Letter} a new equilibrium is proposed, motivated by the
qualitative argument of the previous paragraph.  In this equilibrium, the
disk remains geometrically thin, but most
of the disk mass is found in very dense clumps, with only a small fraction
left behind to form a volume-filling substrate through which the clumps move.
The multiple requirements of dynamical and thermal equilibrium strongly
constrain the allowed parameters of such an arrangement; as
shown in \S 3, these compare well with those inferred empirically.

    Others have proposed that accretion disks around
black holes might contain numerous small clumps, but their efforts have
all focussed on the internal state of the clumps and how they reradiate
the energy they absorb (Guilbert \& Rees 1988;
Celotti, Fabian \& Rees 1992; Kuncic, Blackman \& Rees 1996;
Kuncic, Celotti \& Rees 1997).  Here we examine not just the thermal properties
of the clumps, but the overall equilibrium of the accretion disk.

\section{Equilibrium Solution}

   The existence of this new equilibrium depends on one key assumption: that
whatever process forms the clumps, it leaves them magnetically connected
to the volume-filling plasma.  This would be the result if, for example,
they are created by thermal instabilities, or a ``photon bubble" instability
(Gammie 1998).
The specific structure of these connections is unimportant.  All that
is required is for most of every clump's mass to be attached
to a field line that ultimately makes its way out into the external medium.

  With this assumption, we begin defining this equilibrium
by applying the requirement of angular momentum
conservation.  Here, and throughout this
{\it Letter}, we deal exclusively with vertically-integrated
(or vertically-averaged) quantitites.  The dominant angular momentum
transport mechanism is likely to be clump-clump collisions.  Magnetic
torques may also play a role, but are likely to be smaller.  This is
because the $r-\phi$ component of the magnetic stress cannot exceed
$B^2/8\pi$, yet the magnetic energy density is at most comparable
to the gas pressure, which, as we estimate below, is probably well
below the kinetic energy density of the clumps.

  When (as can be expected here) the collision frequency between clumps is comparable to or smaller than the orbital frequency, the effective
viscosity due to collisions is reduced (Goldreich \& Tremaine 1978).  Taking
the unit of stress to be the momentum flux density of clumps $P$, we
describe viscous transfer of angular momentum by clump collisions
in terms of an effective ``$\alpha$"
parameter (as in Shakura \& Sunyaev 1973): $\alpha_C \simeq \min(C,C^{-1})$,
where $C$ is the covering fraction of clumps along a vertical line
of sight.  In the same units, any other $r-\phi$ stresses (e.g. magnetic)
add an amount
$\alpha_M$.  The angular momentum conservation equation may then be written
as a constraint on the total (i.e. hot phase plus cold) Thomson
optical depth measured from the midplane to the surface:
\begin{equation}
\tau_T = (9/5)(\bar{m}/m_e) {\dot m R_T \over (\alpha_C + \alpha_M)
{\cal M}^2
 \Theta x^{3/2}} = 64 {R_T \over (\alpha_C + \alpha_M){\cal M}^2}
\mdotscale \Thetascale^{-1} \xscale^{-3/2}.
\end{equation}
Here $\bar{m}/m_e$ is the ratio of the mean mass per particle to the
electron mass, $\dot m$ is the accretion rate in Eddington units (for
unit efficiency), $\Theta$ is the hot phase temperature in units of $m_e c^2/k$,
${\cal M}$ is the velocity dispersion of the clumps
in units of the hot phase sound speed, $x = r c^2/GM$, and $R_T$ is the
relativistic correction factor to the integrated stress (Novikov
\& Thorne 1973).  Note that at $x = 10$, $R_T \simeq 0.1$ -- 0.5,
depending on the black hole spin.  We expect, therefore, that if (as
estimated below) the other fiducial factors are all order unity, in
sub-Eddington accretion $\tau_T$ is never more than several tens,
and could be rather less.

    The origin of the clumps' random motions lies in their magnetic connections.
Consider a flux tube that passes
through two clumps initially at the same azimuthal angle, but
slightly different radii.  As the inner one moves ahead of the outer one,
the field energy grows as the tube lengthens.  The associated force
transfers angular momentum outward, and the clumps' radial separation grows.
If the field is initially weak (in the sense that the magnetic force
between a pair of clumps is smaller than the central mass's gravity),
this is an unstable process, very similar to the continuum fluid
magneto-rotational instability whose importance to accretion disk evolution was
pointed out by Balbus \& Hawley (1991).  Numerical integration of the
equations of motion demonstrates that, if left uninterrupted, the end
result is to give clumps energies of random motion comparable to the initial
difference in gravitational potential energy between
linked clumps.  At the end of this section we will estimate the typical
separation of clumps to show that the expected
random speeds are then of order the hot phase sound speed.
The field line stretching adds energy to the magnetic field,
which may saturate at a level that is roughly in equipartition with the gas
pressure.

   Next consider energy conservation in the hot phase.  Several
heating mechanisms act, all having roughly
constant rate per unit volume.   Field lines running through a clump and the
adjacent hot gas are pulled along by the clump, and exert a drag force
on the external plasma.   The energy of the random motions so induced is
eventually dissipated into heat.  The magnetic field can be dissipated by
forced reconnection (when clumps attempt to pull field lines across one
another) and other mechanisms.  Because the dissipation
draws energy from the random motions of the clumps {\it relative} to the
local mean velocity, it heats the hot, low-density
phase, but exerts no torque.

    In disk-stress units, the hot phase heating rate is $\alpha_H P \Omega$,
where $\Omega$ is the orbital frequency.  We expect that
$\alpha_H \sim C {\cal M} (\tau_h/\tau_T){\cal R}$, where $\tau_h$ is
the (half) Compton optical depth of the hot phase, and ${\cal R}$ is
the ratio between the effective drag cross section of a clump and its
geometrical cross section.  ${\cal R}$ can be rather more than unity, but
it cannot be less than $\sim 1$.  Because the other parameters in the
estimate for
$\alpha_H$ are all regulated to be $\sim 1$ (see below), and generally
$\tau_T < 30$, although $\alpha_H$ could be smaller than
$\alpha_C$, it cannot be too small. 

    The hot phase is cooled by inverse Compton scattering.  Pietrini \& Krolik (1995) showed that when $\tau_h \sim 1$, the thermal balance of a plasma cooled
by inverse Compton scattering may
be described approximately by an expression, which, in this context,
becomes
\begin{equation}
\Thetascale \tau_h = \left(C + \alpha_C/\alpha_H\right)^{-1/4}.
\end{equation}

    The clumps are heated both by dissipative collisions and by absorbing
X-rays radiated by the hot phase.  If they reradiate thermally and $C < 1$,
their outer surface temperature is
\begin{equation}
T_s = 6.2 \times 10^6
\left({\alpha_C/C + \alpha_H \over \alpha_C + \alpha_H}\right)^{1/4}
\mdotscale^{1/4} \xscale^{-3/4} m^{-1/4} R_{R}^{1/4} \hbox{ K} ,
\end{equation}
where $m = M/M_{\odot}$ and $R_R$ is the relativistic correction factor
for the dissipation rate per unit area (Novikov \& Thorne 1973).
If, as is likely, $\alpha_C/C > \alpha_H$, so that most of the dissipation
associated with $\alpha_C$ occurs deep inside the clumps,
the temperature at their centers is larger than $T_s$ by a factor
$\sim (1 + \tau_c)^{1/4}$,
where $\tau_c$ is the Rosseland mean optical depth through a clump.

    In contrast to, e.g. Kuncic et al. (1997), who advocate magnetic
confinement of the clumps, we suppose that the clumps are magnetically
linked to the hot phase.  Consequently, motion along field lines is
unimpeded by magnetic forces, and clump survival
requires pressure balance along field lines.  Although it is true,
as they argue, that maintenance of a smooth internal pressure distribution
matched to the external pressure requires clumps small enough
for sound waves to cross during a dynamical time, the picture proposed
here does not depend on the clumps maintaining a smooth internal pressure
distribution.  All that is really necessary is that the pressure inside
a clump does not vary by
large factors.  If ${\cal M} < 1$, the variations in the external pressure
at a clump's edge are only of order unity, so there is no
reason to impose such a strict upper bound on the clump size.

    The gas pressure in the hot phase is generally somewhat greater than
the radiation pressure.  Approximate pressure balance with the clumps then
implies a gas density in the clumps   
\begin{equation}
n_{cl} \simeq 3 \times 10^{21} \tau_h \left({\alpha_C/C + \alpha_H \over
\alpha_C + \alpha_H}\right)^{-1/4} \mdotscale^{-1/4} \Thetascale^{1/2}
\xscale^{-3/4} m^{-3/4} R_{z}^{1/2} R_R^{-1/4} \hbox{ cm$^{-3}$},
\end{equation}
where $R_z$ is the relativistic correction factor for the vertical
gravity (Abramowicz et al. 1997).  At such high densities, the
approximation of thermal radiation (equation 3)
should be reasonably valid, even when (as in AGN) $m \sim 10^8$.

    Thermal equilibrium in the presence of heat conduction (which in this
context must flow along field lines) constrains both the size of the clumps and
the external pressure.  As shown by McKee \& Begelman (1990), when
conduction is in the classical regime, the relative motion between the
clumps and the hotter gas around it is subsonic, and the geometry is
plane-parallel, there is a unique external pressure
at which there is neither condensation nor evaporation.
Plane-parallel flow occurs in one of two ways: if the clumps are larger than the
length scale on which the temperature varies, or if the magnetic field lines are
parallel.  With the assumption that pathlength integrated along field lines is not grossly different from straight line pathlength, the characteristic length
for the temperature gradient at
which radiative cooling becomes comparable to thermal conduction is the
Field length $\lambda_F \equiv n_e^{-1}(\kappa T/\Lambda)^{1/2}$,
where $n_e$ is the electron density in the conducting gas, $\kappa$ is
its conductivity, and $\Lambda$ is its cooling function.

    The plane-parallel condition is met if the clumps are at least as large as
$\lambda_F$.  The unique equilibrium pressure is defined by the condition
that radiative heating and cooling exactly balance when integrated
across the regions of intermediate temperature.  When there is no net
evaporation or condensation, the gas in the interface is static
({\it modulo} any relative bulk motion).
Its pressure is therefore nearly constant.  Since
the rate of Compton cooling per unit volume is proportional to
the electron pressure so long as the electrons are sub-relativistic,
it is independent of temperature within the conductive interface.  Consequently,
if the heating rate is fixed per unit volume, the difference between the
heating rate and the Compton cooling rate is also constant throughout the
interface.  This fact means that the relevant cooling rate
for determining $\lambda_F$ is the (small) part due to bremsstrahlung.

    Relative to the disk scale height, $\lambda_F$ is:
\begin{equation}
\lambda_F/h = 0.26 \tau_h^{-1} \Thetascale^{3/2} = 0.26 \tau_h^{-5/2}
\left(C + \alpha_C/\alpha_H\right)^{-3/8},
\end{equation}
where the second expression has used equation 2 to fix the temperature
in terms of the optical depth.  The numerical values in equation 5 are
computed assuming the Spitzer conductivity,
$\kappa = 5.6 \times 10^{-7}$~erg~cm~s$^{-1}$~K$^{-1}$.
Equation 5 shows that to achieve evaporative balance,
the clumps must be an interesting fraction of a scale height across
(at least in the plane perpendicular to the local magnetic field
direction).  This equation also places an implicit lower bound on $\tau_h$.
The reason is that if $\tau_h$ were so small that $\lambda_F/h > 1$, clouds
large enough to have plane-parallel conductive interfaces would also be
so large that the surrounding pressure could not everywhere be equal to the
equilibrium pressure.

     An upper bound on $\tau_h$ also follows.  The differential equation of
thermal balance allowing
for dissipative heating in the hot phase, Compton cooling, bremsstrahlung, and
heat conduction may be solved
subject to the boundary condition that the temperature gradient goes to zero at
large distance from a clump.  The condition of evaporative balance then
reduces to
\begin{equation}
H_{net} \int_{0}^{1} \, dt \, t^{5/2} \left[(4/7)H_{net} (1 - t^{7/2}) - 
(1 - t^2)\right]^{-1/2} = \int_{0}^{1} \, dt \, 
t \left[(4/7)H_{net} (1 - t^{7/2}) -  (1 - t^2)\right]^{-1/2},
\end{equation}
where $H_{net}$ is the ratio of net heating rate (i.e. after subtracting
the Compton cooling rate) to the bremsstrahlung cooling rate, evaluated
far from any clump.   The solution to equation 6 is $H_{net} \simeq 1.765$.
Rewriting $H_{net}$ in terms of $\tau_h$ then gives:
\begin{equation}
0.15 \mdotscale^{-1} \xscale^{3/2} R_{z}^{1/2} R_{R}^{-1} \tau_h^{2} + 
\left(C + \alpha_C/\alpha_H\right)^{1/4} \Thetascale \tau_h - \left(1 +
\alpha_C/\alpha_H\right)^{-1} = 0.
\end{equation}
When Compton cooling dominates, $\Theta$ adjusts to permit an equilibrium for
any value of $\tau_h$.  However, if $\tau_h$ is too large, $\Theta$ falls
to the point that bremsstrahlung may contribute significantly.  The
critical optical depth is
\begin{equation}
\tau_{h,max} \simeq 2.5 \left(1 + \alpha_C/\alpha_H\right)^{-1/2}
\mdotscale^{1/2} \xscale^{-3/4} R_{R}^{1/2} R_{z}^{-1/4} .
\end{equation}
When $\tau_h$ becomes this large, evaporative balance fixes $\tau_h = 
\tau_{h,max}$.  Thus, bounded by the requirement that $\lambda_F < h$
and by equation 8, $\tau_h$ must always be $\sim 1$.

    The covering factor may be found from the geometric relation
$C \simeq (T_s/T_h)(\tau_T/\tau_h)(h/\lambda_F)$ if the typical clump
temperature is $\simeq T_s$ and the typical clump size
is $\sim \lambda_F$. Evaluating this expression and solving for $C$ (assuming
$C < 1$, we find
\begin{eqnarray}
C \simeq 0.30 {(1 + \alpha_H)^{9/11} \over (1 + \alpha_H/C)^{2/11}
(1 + \alpha_M/C)^{8/11}} &{\cal M}^{-16/11}
\left({\alpha_H \over 0.1}\right)^{-7/11} m_{8}^{-2/11} \mdotscale^{10/11}
\nonumber \\
 &\times {\tau_h}^{28/11} \xscale^{-18/11} R_{R}^{2/11} R_{T}^{8/11}, 
\end{eqnarray}
where we have assumed that $\alpha_{M,H} < \alpha_C < 1$.  In this
equation we
scaled the central mass to $10^8 M_{\odot}$ because $C$ rises toward
unity as the central mass declines toward the solar scale.  When $m \sim 1$,
$C < 1$ is no longer a good approximation, although $C$ does not exceed
unity because of its strong dependence on $\tau_h$, whose allowed
range shifts slowly downward with increasing $C$ (equations 5 and 8).

    Finally, we estimate $a$, the mean separation between clumps, in terms of
$C$:
\begin{equation}
a/h \simeq 0.4 C^{-7/12}  (1 + \alpha_H)^{-1/4} \tau_h^{-5/6} \left({\alpha_H
\over 0.1}\right)^{1/4} .
\end{equation}
Thus, $a/h$ is regulated to be $\sim 1$.  The difference in potential energy between neighboring clumps is then great enough, that, given weak magnetic connections, they are stirred up to ${\cal M} \sim 1$.

\section{Discussion}

   So long as the hot phase cools predominantly by inverse Compton
scattering, it should be thermally stable, for the Compton cooling rate
automatically rises with the energy density of photons available.   On the
other hand, should $\tau_h$ rise to $\tau_{h,max}$, the optical depth at
which bremsstrahlung becomes significant, the hot gas would become unstable
to thermal perturbations with wavelengths short compared to a scale height.
The result would be the creation of more cool clumps,
so that $\tau_h$ would be brought back into the permitted range.
Thus, $\tau_{h,max}$ is truly an upper bound on the
optical depth that may be found in the hot phase.  The equilibrium is
viscously stable because combining equations
1, 2, and 9 shows that $\dot m \propto \tau_T^{22/7}$.

   At our fiducial values, there is only a factor of several
between the largest and smallest possible $\tau_h$.  However,
the dependence of $\tau_{h,max}/\tau_{h,min}$ on $C$ is such that the
net scaling with both $\dot m$ and $x$ is extremely slow: $\tau_{h,max}/
\tau_{h,min} \propto {\dot m}^{2/11} x^{-39/220}$.  Consequently, $\tau_h$
is constrained to within factors of a few
for any values of $\dot m$ and $x$ for which
the canonical disk equilibrium indicates radiation pressure dominance.
This tight constraint also ensures that $p_g/P \sim \alpha (m_e/\mu_e) \tau_h
x^{3/2}/(\dot m R_T) \ll 1$.

    Finally, we compare the observables predicted by this equilibrium
to those actually seen.  Although the estimates made in this {\it Letter} are
highly approximate, each of them can be refined with more detailed
calculation.  The most important prediction of this equilibrium is the
simultaneous presence of both a thermal and a ``coronal"
component in the spectrum.  More specifically, it predicts, as
is seen in real black hole systems (Zdziarski et al. 1996),
that $\tau_h \sim 1$ and $\Theta \sim 0.1$, independent of the
central mass.  It also predicts that the covering fraction of the clumps
depends weakly on $m$, although it rises with increasing $\dot m$ and falls outward.  In the inner rings, $C \sim 0.1$ for AGN, but is closer to
$\sim 1$ for stellar black holes.  In AGN, each clump is very optically
thick to Compton scattering, so they are individually (and collectively)
effective absorbers of soft X-rays, and reflectors of hard X-rays.  In
disks accreting onto stellar black holes, the individual clump optical
depth may not be so great, so that the Compton reflection feature may
be weakened.  $C$ is in general less than
unity, so we would not expect (as we do not generally see) strong soft X-ray
absorption.  The slope of the Comptonized power-law depends on
$(C + \alpha_C/\alpha_H)^{1/4}$
(Pietrini \& Krolik 1995).  If $\alpha_H$ is not too much smaller than
$\alpha_C$, the slope predicted by this relation is consistent with
observations (i.e. $F_\nu \propto \nu^{-0.9}$ in AGN: Mushotzky, Done,
\& Pounds 1993; in Galactic black hole systems, the power-law index
ranges from $\simeq 0.3$ to $\simeq 1.5$: Tanaka 1989; Ballet et al. 1994;
Gil'fanov et al. 1994).
The ratio of hard X-ray luminosity to thermal emission (ultraviolet in AGN,
soft X-ray in stellar mass black hole systems) is given by $\alpha_H/(\alpha_C
+ C\alpha_H)$.

\acknowledgments

   I thank Eric Agol, Steve Balbus, and Ramesh Narayan for helpful
conversations.  This work was partially supported by NASA Grant
NAG5-3929 and NSF Grant AST-9616922.

\end{document}